\documentclass[doublecol]{epl2} 

\title{Large deviations and heterogeneities in a driven kinetically constrained model}

\author{F. Turci \and E. Pitard}

\institute{Laboratoire des Collo\"{\i}des,
 Verres et Nanomateriaux \\
  Universit\'e Montpellier II and CNRS, 34095 Montpellier, France}


\abstract{Kinetically Constrained Models (KCMs) have been widely studied in the context of glassy dynamics,
 focusing on the influence of dynamical constraints on the slowing down of the dynamics of a
macroscopic system. In these models, it has been shown using
 the thermodynamic formalism for histories, that there is a coexistence between an active and an inactive
phase. This coexistence can be  described by a first-order transition, and a
related discontinuity in the derivative of the large deviation function for the activity. We show that adding a
driving field to a KCM model does not destroy this first-order transition for the activity. Moreover, 
a singularity is also found in the large deviation function of the current at large fields.
We relate for the first time this property to microscopic structures, in particular
the heterogeneous, intermittent dynamics of the particles, transient
shear-banding and blocking walls. We describe both the shear-thinning and the shear-thickening 
regimes, and find that the behaviour of the current is well
reproduced by a simple model.
}

\begin{document}

\maketitle

\section{Introduction}

Disordered materials, such as gels, glasses, complex fluids or granular materials, exhibit slow
relaxation and heterogeneity, both in space and time. In particular, spatial dynamical
heterogeneities, i.e the coexistence of mobile particles and blocked particles,
 have been widely studied \cite{DH} and have benefited from the
results of KCMs (Kinetically Constrained Models), which by definition rely on specific dynamical
rules depending on the mobility of the particles \cite{Ritort-Sollich}.

When driven by an external force, these complex systems exhibit specific spatial features, such
as shear-banding \cite{LBTG-FIELDING}. 
Another complication arises  \cite{shear-thickening}  with shear-thickening, where
viscosity increases when  the driving force increases \cite{larson}.
In this paper, we study a KCM driven by an external field, using both global observables (large
deviation functions) and a microscopic description. This model, first introduced by Sellitto
\cite{sellitto}, has the advantage of displaying both shear-thinning and shear-thickening regimes: 
at small field, the current of particles is proportional to the field, whereas at large fields,
the system has a negative differential resistance, i.e the current decreases.
Similar effects have also been observed in electronic systems \cite{Shklovskii} where
negative differential conductivity can occur under strong electric fields.

In this context we have studied the thermodynamics of histories in the
 approach developed by Ruelle \cite{Ruelle, Touchette} 
for deterministic dynamical systems. This formalism can be extended to  Markov chains  
\cite{LAW}, and has been already developed for KCMs
  \cite{GJLPW}. In these latter studies, where the observable is the activity, one is able
to probe trajectories in phase space with either finite, or almost zero activity, according to a
"chaoticity temperature", $s$, which is introduced in the formalism. 
Contrarily to systems without
kinetic constraints, there is a first-order transition at $s=0$ in the activity,
 which translates
into a discontinuity of the derivative of the associated large deviation function.
 This
transition is the global signature of the observation of dynamical heterogeneity. 
In the case of
driven KCMs, we will focus on the large-deviation functions 
for both the activity and the current. We will also present a microscopic analysis of the
structures involved in the dynamics, and relate them to global quantities.

\section{Ruelle's formalism}
 This formalism yields information about the fluctuations
of temporal trajectories in configuration space, while the usual canonical thermodynamics approach yields
information about the fluctuations associated with configurations.
One defines a dynamical partition function, with a chaoticity temperature that is conjugated to the
observable. Hence, if $O$ is a time-extensive observable, $Z_O(s,t)=\left< e^{-sO}\right>$. 
The brackets
mean that one is averaging over all possible histories between $0$ and $t$.
More precisely,
$Z_O(s,t)=\sum_{histories} {\rm Prob(history)} e^{-sO({\rm history})}$,
 where a history is the time series of
configurations $\left\{C_{1}, C_{2},\dots,C_{t}\right\}$. 
      In the long time limit, the partition function behaves like
       $$\lim_{t\rightarrow \infty}Z_O(s,t)
       \propto \exp\left\{t\psi_{O} (s)\right\}$$ 
where $\psi_{O}$
        is the large deviation function associated with the 
	observable $O$. 
In practice, one is also interested by the mean of the observable in $s$-space, i.e, where the
trajectories are biased by the factor $e^{-sO}$. This quantity is related to the large deviation
function through the relation, in the large time limit:
$<O(s,t)>/t=-\psi'_O(s)$.

\section{The model}

In this paper, we study a two dimensional asymmetric simple exclusion process (ASEP) on a 
square lattice with dynamic constraints \cite{sellitto}. 
One can also view the model as the KA (Kob-Andersen) model
 in two dimensions -an example of KCM- \cite{Kob-Andersen},
 driven by an external
field. The dynamical rule is such that a particle can move to a neighboring site only if it has
at least two empty neighbours, before and after the move. Such a constraint mimics the fact that a
particle in a dense disordered material can only move if it has enough free volume. Such a model
always has a non-vanishing diffusion coefficient except for a density equal to $1$ \cite{Toni}.
The motion of the particles is biased in one direction of the lattice by an external field 
$\vec E$,
such that
 the probability for a particle to  move in a certain direction 
 is $p=\min\left\{1,\exp{\left[-\vec{E}\cdot\vec{r} \right]}\right\}$, where
  $\vec{r}$ indicates the displacement vector. The boundary conditions are periodic in both
  directions. In \cite{sellitto}, the current was measured as a function of density $\rho$ and the field,
  showing several distinct behaviours.
For densities below a threshold 
value (approximately $\rho_c\approx 0.79$) the system shows a monotonic growth 
of the current as a function of the external field (like in the non constrained ASEP model), which
eventually saturates to a maximum current value.
For $\rho\geq \rho_c$, the current has two different regimes: for $E$ small, a first ohmic
   regime (also referred to as a "shear thinning" regime) 
  where the current grows proportionally to the field; and for $E$ large, a 
  negative differential resistance regime (also called "shear thickening" regime) where the
   current decreases as the field increases. Such a behaviour has also been observed 
    in other similar models of glasses
    \cite{JKGC}. 
    In this paper, we are mainly interested in the origin
     of the non-monotonic  behaviour of the current as a function of the field. As a consequence
       we will 
     consider only densities above the critical density $\rho_c$. 
\begin{figure}[t]
\begin{center}
\includegraphics[width=\columnwidth]{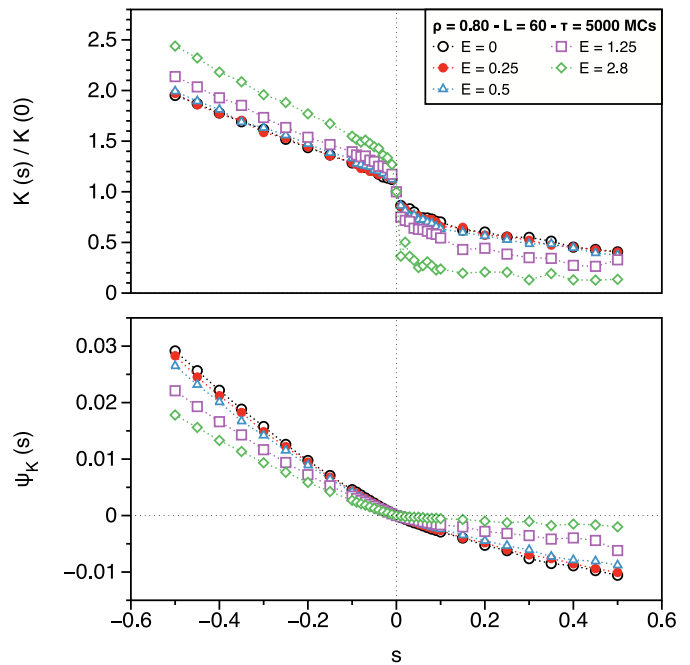}
\caption{The mean activity  $K(s)$ in the modified dynamics, normalised by its value in $s=0$ (top), 
and its large deviation function $\psi_{K}(s)$ (bottom) for a system of 
 size $L^2=60^2$, at high density $\rho=0.80$. For the simulation, we have 
considered 300 clones of the system evolving for a total time of $\tau=5000$ MCs. 
For any value of the external field there is a discontinuity in the derivative of 
$\psi_{K}$ which corresponds to two different regimes in $K(s)/t$:  an active one 
($s<0$) and an inactive one ($s>0$).}
\label{activity}
\end{center}
\end{figure}

In the following, we will consider two different time extensive
observables, namely the activity $K$ and the integrated current $Q$.
The activity $K(t)$ is defined as the total number of moves
 from time 0 to time $t$. The integrated current $Q(t)= \int_{0}^{t}J(t')dt'$
 is defined as the number
  of moves in the direction of the field from time 0 to time $t$.
   The current calculated in \cite{sellitto} is simply
  $Q(t)/t$.

In order to simulate the biased dynamics (the ``s-dynamics'') and to 
compute the values of $K(s)$,  $Q(s)$  and their large deviations 
functions $\psi_{K}(s)$ and  $\psi_{Q}(s)$, we have used the discrete 
time cloning algorithm proposed in reference \cite{Giardina}, which was proved to be powerful
enough for numerical studies of KCMs.

\begin{figure}[t]
\begin{center}
\includegraphics[width=\columnwidth]{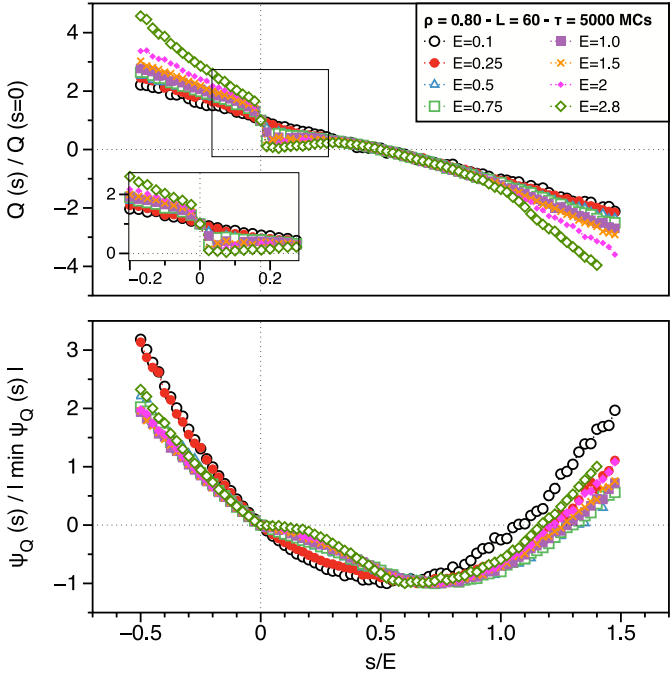}
\caption{The mean integrated current $Q(s)$ in the modified dynamics normalized by the
 value $Q(s=0)$(top) and its large deviation function $\psi_{Q}(s)$ (bottom) 
plotted with a proper rescaling. The system  has a size of $L^2=60^2$
 sites and a density $\rho=0.8$. The
 maximum current phase is attained for a field $E^* \simeq 2$. 
 Here we have considered 300 clones of the system evolving for $\tau=5000$ MCs.
 The increase of the discontinuity of $Q(s)$ at $s=0$ is emphasized in the inset.}
\label{current}
\end{center}
\end{figure}

\section{Large deviations computations}
The large deviation function for the activity has already been studied 
 for different KCMs without driving such as
the Fredrickson-Andersen Model, the triangular lattice gas model, 
the Kob-Andersen model \cite{GJLPW}.
A first-order phase transition in $K$ was observed at $s=0$. 
Such a transition is the global signature for
the coexistence at $s=0$ (which corresponds to the unbiased, physical system)
of histories that have 
positive values of the activity (typical of $s<0$), and histories that have 
sub-extensive activity (typical of $s>0$). It is a translation in phase space of the dynamical
heterogeneities one finds when looking at configurations.
 For non-driven  models the activity's phase 
transition is directly linked to the presence of kinetic constraints, 
since it disappears as soon as the constraints are removed. 
In our driven model,  we find that this first-order transition still exists for the activity
(see Fig. \ref{activity}), and is by no way smeared out by the driving, whether 
the system is in the positive  or in the negative differential resistance regime. The discontinuity observed at $s=0$
increases as $E$ gets larger. We checked that for all fields, the transition
exists at the thermodynamic limit: we performed a finite-size study showing that the discontinuity
is more and more visible as the size of the system increases. In order to do this study accurately,
we needed more and more clones and longer simulation times for larger system sizes, in order for the
cloning algorithm to converge.

\begin{figure}[htbp]
\begin{center}
\includegraphics[width=\columnwidth]{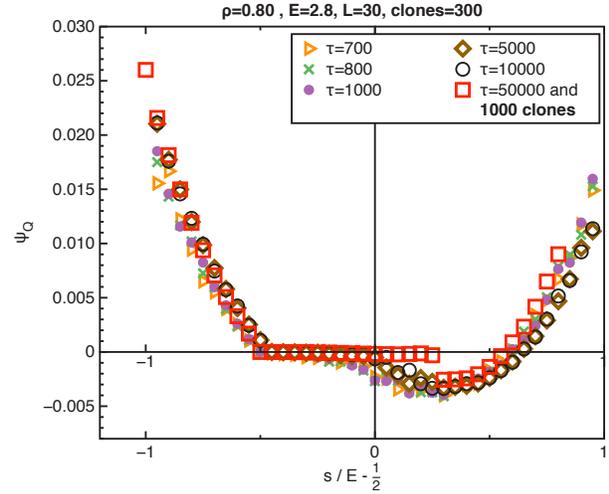}
\caption{The  large deviation function $\psi_{Q}(s)$  
for a system   size of $L^2=30^2$
 sites and a density $\rho=0.8$. 
 Here we have considered 300 clones  evolving for different times
 $\tau$. We also show the result for 1000 clones and a very long simulation time, suggesting the
 asymptotic symmetric shape, and a clear discontinuity of the derivative
 at $s=0$.}
\label{currentlongtimesandclones}
\end{center}
\end{figure}

We considered then the 
integrated current, flowing 
 in the direction of the external field. 
It  is linked to the activity since 
 it is the difference between the activity in the positive direction 
 of the field and the one in the negative direction. In the case of a 
 non-constrained asymmetric exclusion process in one dimension, 
 it has been shown \cite{Bodineau-Derrida} 
 that the current's large deviation function is a regular convex function with a symmetry 
 around $E/2$ due to the fluctuation theorem \cite{Lebowitz-Spohn}. 
 This means that no discontinuity exists for
  any value of $s$ for the derivative of $\psi_{Q}(s)$ in the non
   constrained case, or, in other terms, $Q(s)$ is a continuous
    function. We checked this in the case of the
    two dimensional ASEP. Moreover, a perfect collapse of the $Q(s)/t$ and $\psi_Q(s)$ curves can be found
    when rescaling $s$ by $s/E$ and the ordinates according to the scaling used in Fig. \ref{current}.
    
If we introduce the dynamical constraints we find (see Fig. \ref{current}) 
 corrections to this perfect scaling.   For small fields no discontinuity can be found
 (even if one uses more clones and longer times in the simulation); the large deviation
     function $\psi_{Q}(s)$ has the same shape we should expect for a non constrained model
     (2d-ASEP).
      Conversely, if we increase the field, a gap between the left and 
      the right derivatives of $\psi_{Q}(s)$ is generated, 
      and the size of the gap increases continuously
as we increase the field. 
      In the limit of large system sizes, the discontinuity in
       the derivative of $\psi_{Q}(s)$ corresponds to a step 
       discontinuity in the biased current $Q(s)$; we also performed this finite-size study, and
       found that the discontinuity is sharper and sharper as the size increases. 
 Here, the convergence of the cloning algorithm at fixed $L$ is rather slow. Figure
\ref{currentlongtimesandclones} illustrates this point: as time and the number of clones increase,
the transition is more and more visible. This figure also suggests that $\psi_Q(s)$ should
asymptotically converge  to a symmetric shape with a central part equal to zero; this is actually
what is predicted by the fluctuation theorem \cite{Lebowitz-Spohn} and was found in a mean-field
model \cite{Speck-Garrahan}. This asymptotic curve could not be obtained here, due to numerical
limitations: this difficulty was already raised by Sellitto \cite{sellitto} who also found
deviations from the fluctuation theorem, even at very large times.

\begin{figure*}[htbp!]
\begin{center}
\includegraphics[height=144pt]{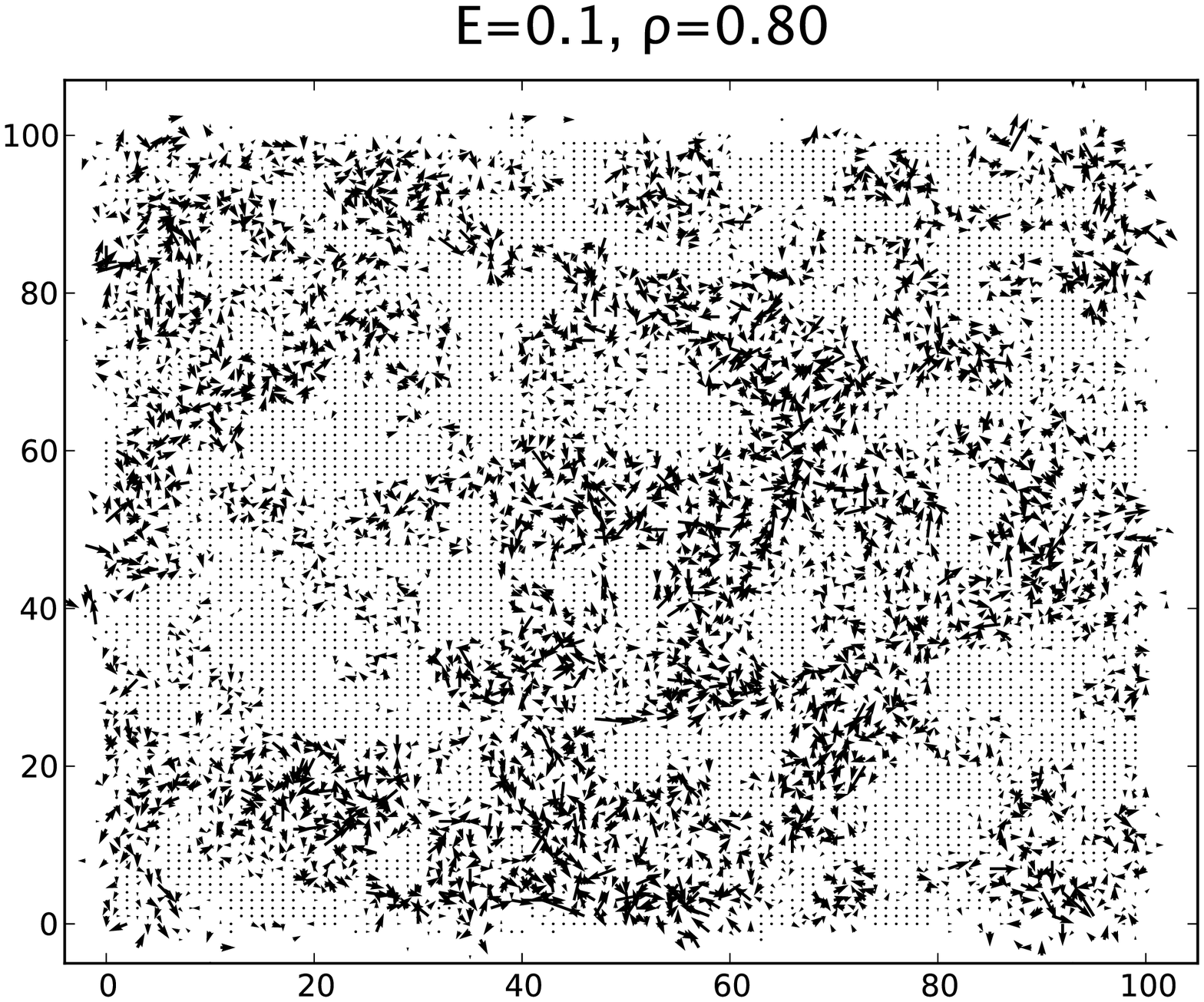}
\includegraphics[height=144pt]{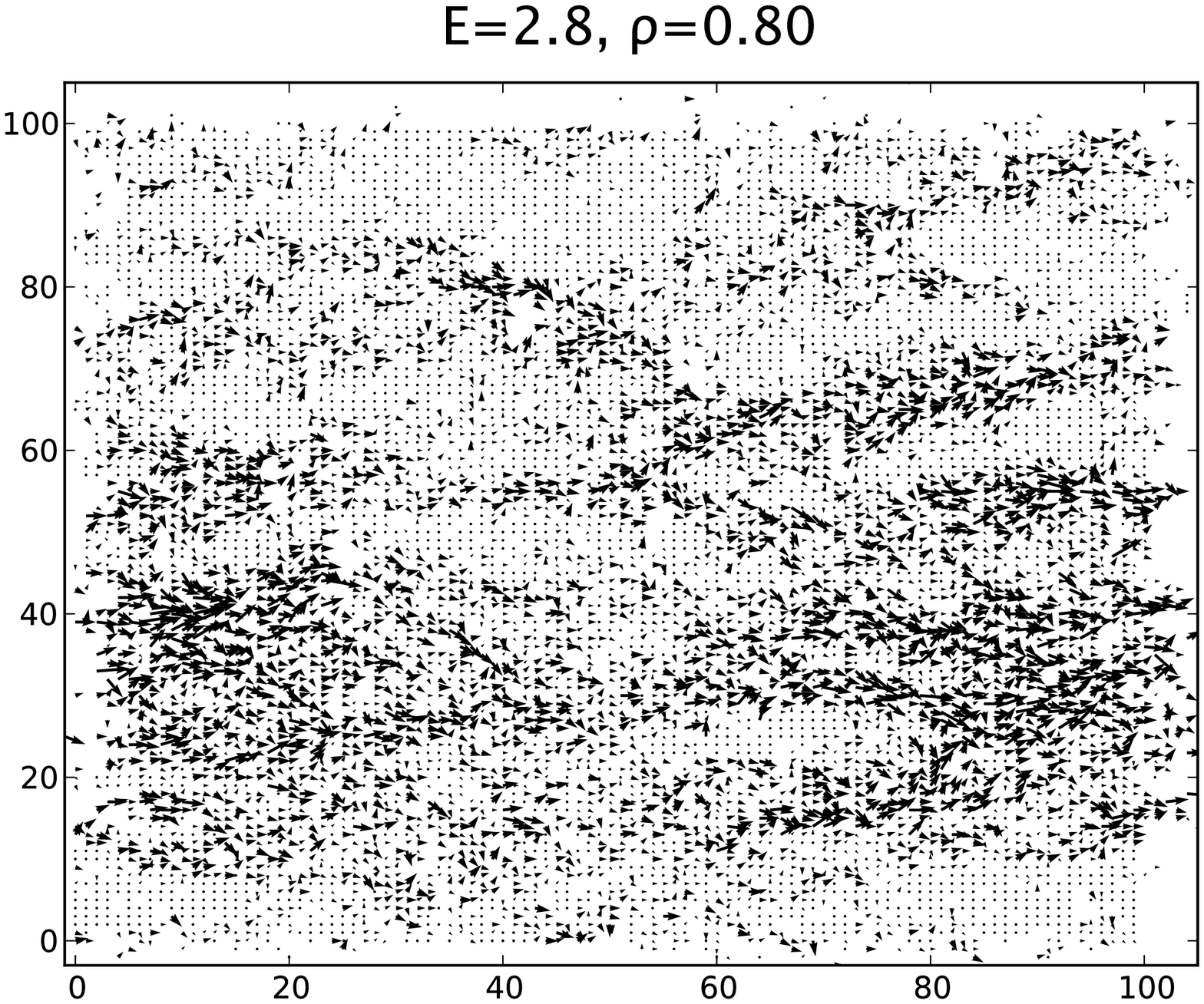}
\includegraphics[height=139pt]{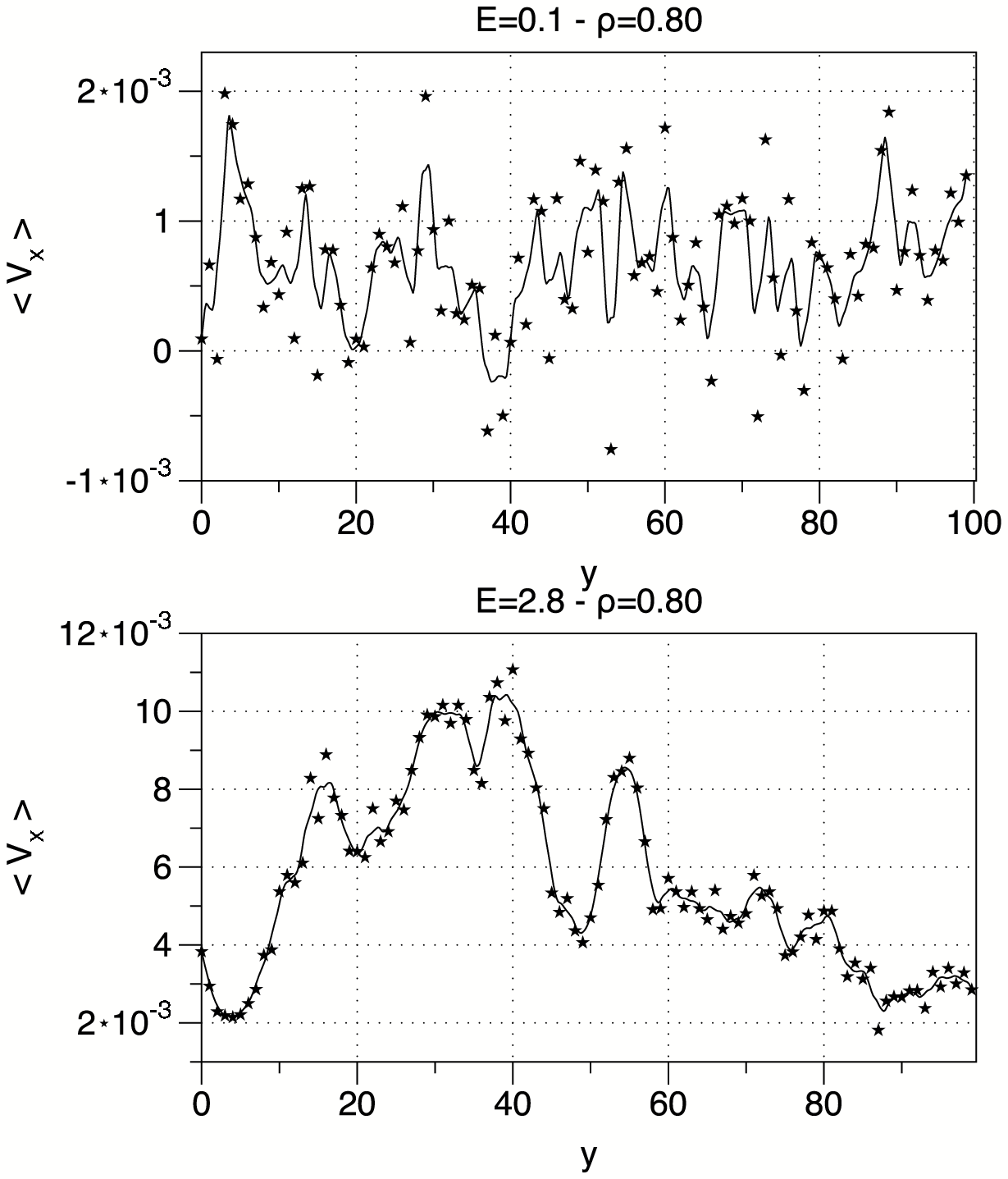}
\caption{On the left and middle: velocity field plots 
for two values of the external field, corresponding to the same value of the average current. The diffusive homogeneous behaviour at small fields can be
contrasted with the localisation of transport in bands for large fields.
On the right: the mean longitudinal velocity
 versus the direction transversal to the external field, for the same
  two values of the driving force.}
\label{quivers}
\end{center}
\end{figure*}

\section{Microscopic analysis}
Physically, this transition in $Q(s)$ and 
$\psi_Q(s)$ at $s=0$ corresponds to a coexistence
      between histories that  have a 
       positive current ($Q(s)$ is positive for $s<0$) from 
       histories that have a subextensive current (for $s\rightarrow 0^+$ we have no current, and
       nearly all particles are blocked). 

In order to better investigate 
the emergence of the negative differential resistance behaviour, we have 
analyzed some microscopic aspects of the dynamics. 
These microscopic observations are also useful in order to
understand the behaviour of the large-deviation functions described above, 
in the regimes of small and large forcing.

\begin{figure}[htbp]
\begin{center}
\includegraphics[width=\columnwidth]{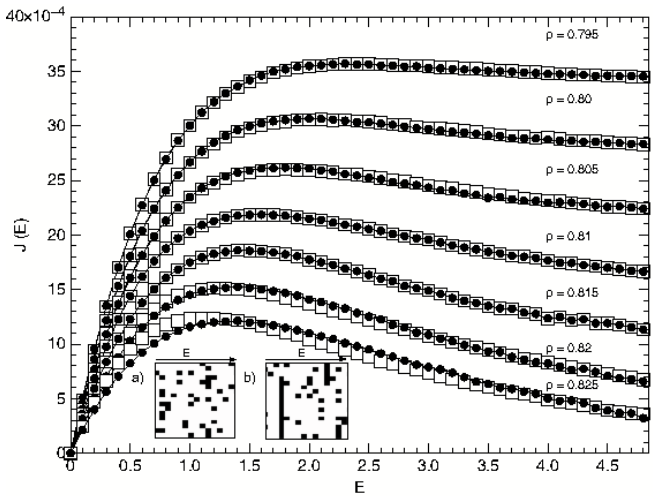}
\caption{Simulated flow curves (white squares) fitted by the phenomenological model
$J(E)=A (1-e^{-E}) (1- \alpha <w>)$ (black dots).Densities range from 0.795 to 0.825, $L=100$, and $<w>$
is determined independently from the simulation. Below are shown two fractions of snapshots
 of configurations taken
at $\rho=0.82$, a) E=0, b) E=5, illustrating the presence of walls. The empty sites are in black, the
particles in white: the empty sites organize into vertical walls at large fields.}
\label{fits}
\end{center}
\end{figure}

Simple observations of the dynamics of the system show that the driving by the field induces
the formation of peculiar void structures, composed of quasi-linear segments of empty spaces, 
in the
direction perpendicular to the field.
 The particles actually accumulate in some dense, blocked regions of the
sample, separated by these "domain walls". These domain walls are hard to remove 
 because of the kinetic constraint; their lifetime is long compared to the microscopic dynamics. In other words, the driving, combined with the dynamical
restrictions, is at the origin of structures that slow down the dynamics. This effect is enhanced
when the field is increased, yielding shear-thickening behaviour.
The domain walls can be directly visualized in the insets of figure \ref{fits} for small and large
fields.

At the level of single particle displacements, the existence of such domains is also at the origin
of different kinds of moves. 
For small fields (and small currents) the behaviour is almost diffusive, 
the particles randomly explore the region around their starting point, 
even coming back and forth on their own path. In the negative resistance regime (large fields) 
 few particles manage to cover very long distances 
 with almost ballistic
trajectories, while others remain caged in the vicinity of their starting point. 
The moving particles actually diffuse in the vertical direction for some time 
before making some  direct jump in the direction of the field. Such jumps actually
correspond to the sudden disappearance of the empty domain walls.
This behaviour is probably at the origin of superdiffusivity \cite{sellitto}, which has
also been observed in granular jammed systems \cite{dauchot-levy}.
Moreover,  it corresponds to a clear intermittent dynamics where the current is locally large for a
short period of time.

We have computed the distribution of  wall sizes $P(w)$ in the stationary regime.
 We have found that
$P(w)$ is well approximated by a simple exponential in a large range of values of $\rho$ and $E$,
allowing to define an average size $<w>(\rho,E)$. This average wall size grows as the density
 or the field grows, and saturates at large values of the external field
\cite{biggerpaper}.We can in fact rationalize the behaviour of the current,
 $J(E)$, by the following phenomenological
argument. Let us assume that one can attribute the slowing down 
of the dynamics by the mere presence of the walls. The
average current  counts the number of particles susceptible
 to move in the direction of the field, and not blocked
by the presence of a wall:
$J(E)=A (1-e^{-E}) (1- p_{blocked})$. 
We assume that the probability to be blocked
 is proportional to the number of  blocked particles 
confined on the left border of a wall. To lowest order in the wall size 
we assume a linear dependence
of $p_{blocked}$ on $<w>(\rho,E)$: $p_{blocked}= \alpha <w>(\rho,E)$,
where $\alpha$ depends in principle on $\rho$ and $E$.
 We found that the curves for $J(E)$ could be very well fitted 
by this formula, taking $\alpha$ independent of $E$ (see Fig. \ref{fits}).
The parameter $\alpha$, which  can be shown in a more
refined study \cite{biggerpaper} to depend on microscopic geometric details, 
such as the number of walls or the average distance between them,
is found to be very weakly dependent
on $\rho$ in the range studied. The assumptions made here can explain 
the fact that the fit actually begins to fail at large $\rho$:  then	
$<w>$ gets larger, 
 $P(w)$ starts to deviate from an exponential
form, showing that the appropriate 
expression for $p_{blocked}$ may change.

At the level of the entire sample, 
we have drawn the velocity field plot  
for the small and large values of the 
driving force, from which one can obtain the velocity profile 
for the longitudinal component of the speed vector (see fig. \ref{quivers}). 
Velocities in these plots are defined over a time lag of the order of the
relaxation time; they also correspond to 
displacement snapshots on this timescale.
These plots  show that at large fields the system gets 
more and more organized, with bands of moving particles coexisting with
static regions blocked by walls, reminiscent of shear banding observed in complex flows.
 This illustrates the first-order
transition for the large deviation of the current at large fields.
The high velocity bands have a macroscopic size and involve the whole 
longitudinal length of the system. At small field, the current is more homogeneous,
which is consistent with the absence of transition in the large-deviation function for the
current.


\section{Conclusion}

We have found that the coexistence  of mobile and blocked trajectories in
configuration space is a conserved feature in KCMs even in the presence of external driving. This is
illustrated quantitatively through the study of the large deviations for the activity and for the
current. In the shear-thinning regime, the 
microscopic behaviour is reminiscent of the dynamical heterogeneities found in KCMs, coming from
dynamical restrictions; there is a transition in the large deviation function for the activity, but not for
the current. In the shear-thickening regime, the field induces a structuration of the
flow, with domain walls separating dense jammed regions,  as well as
transient shear bands. The discontinuity found at large fields in $\psi'_Q(s=0)$
illustrates the dominant role of the blocking walls in the structuration of the displacement field.
We finally note that we have very recently found a preprint by Speck and Garrahan on a
related
subject, where a first-order transition is also found 
for the entropy production for driven KCMs \cite{Speck-Garrahan}.




\acknowledgments
We are grateful to V. Lecomte, F. van Wijland, J. Kurchan, L. Berthier for interesting
discussions, and to M. Sellitto for  correspondance. FT is supported by the French Ministry of Research
and EP by CNRS and PHC no 19404QJ.


\begin{thebibliography}{0}



\bibitem{DH}
  \Name{Sillescu H.}
  \REVIEW{J. Non-Cryst. Solids}{243}{1999}{81}.

  \Name{Ediger M.D.}
  \REVIEW{Annu. Rev. Phys. Chem.}{51}{2000}{99}.

  \Name{Glotzer S.C.}
  \REVIEW{J. Non-Cryst.  Solids}{274}{2000}{342}.

   \Name{Richert R.}
  \REVIEW{J. Phys.  Condens. Matter}{14}{2002}{R703}.

  \Name{Andersen H.C.}
  \REVIEW{Proc. Natl. Acad. Sci. U. S. A.}{102}{2005}{6686}.



\bibitem{Ritort-Sollich}
  \Name{Ritort F.\and Sollich P.}
  \REVIEW{Adv. Phys.}{52}{2003}{219}.
  
  
  
\bibitem{LBTG-FIELDING}
   \Name{Losert W., Bocquet L., Lubensky T.C. \and Gollub J.P.}
  \REVIEW{Phys. Rev. Lett.}{85}{2000}{1428}.
  
    \Name{Fielding F.M., Cates M.E. \and Sollich P.}
  \REVIEW{Soft Matter}{5}{2009}{2378}.
  



\bibitem{shear-thickening} 
 \Name{Liu C.-H. \and Pine D.}
  \REVIEW{Phys. Rev. Lett.}{77}{1996}{2121}.
 

 \Name{Bertrand E., Bibette J. \and Schmitt V.}
  \REVIEW{Phys. Rev. E}{66}{2002}{060401(R)}.

   \Name{Fall A., Huang N., Bertrand F. Ovarlez G.\and Bonn D.}
  \REVIEW{Phys. Rev. Lett.}{100}{2008}{018301}.




\bibitem{larson} 

  \Name{Larson R.G.}
  \Book{The Structure and Rheology of Complex Fluids}
  \Publ{Oxford University Press, Oxford}
  \Year{1999}.


\bibitem{sellitto} 
 \Name{Sellitto M.}
  \REVIEW{Phys. Rev. Lett.}{101}{2008}{048301}.

  
   \Name{Sellito M.}
  \REVIEW{Phys. Rev. E}{80}{2009}{011134}.




\bibitem{Shklovskii} 
 \Name{Levin E.I. \and Shklovskii B.I.}
  \REVIEW{Solid State Commun.}{67}{1988}{233}.

  
    \Name{Nenashev A.V. et al}
  \REVIEW{Phys. Rev. B}{78}{2008}{165207}. 



\bibitem{Ruelle}
  \Name{Ruelle D.}
  \Book{Thermodynamic Formalism}
  \Publ{Addison-Wesley, Reading}
  \Year{1978}.

   \Name{Eckmann J.-P. \and Ruelle D.}
  \REVIEW{Rev. Mod. Phys.}{57}{1985}{617}.

    \Name{Gaspard P.}
  \Book{Chaos, scattering and statistical mechanics}
  \Publ{Cambridge University Press, Cambridge}
  \Year{1998}.



\bibitem{Touchette}
 \Name{Touchette H.}
  \REVIEW{Phys. Rep.}{478}{2009}{1}.



\bibitem{LAW}
 \Name{Lecomte V., Appert-Rolland \and van Wijmand F.}
  \REVIEW{Phys. Rev. Lett.}{95}{2005}{010601}.


\bibitem{GJLPW}
 \Name{Garrahan J.P., Jack R.L., Lecomte V., Pitard E., van Duijvendijk K.
 \and van Wijland F.}
  \REVIEW{Phys. Rev. Lett.}{98}{2007}{195702}.

 \Name{Garrahan J.P., Jack R.L., Lecomte V., Pitard E., van Duijvendijk K.
 \and van Wijland F.}
  \REVIEW{J. Phys. A}{42}{2009}{075007}.

\bibitem{Kob-Andersen} 
 \Name{Kob W.\and Andersen H.C.}
  \REVIEW{Phys. Rev. E}{48}{1993}{4364}.


\bibitem{Toni} 
 \Name{Toninelli C., Biroli G.\and Fisher D.}
  \REVIEW{Phys. Rev. Lett.}{92}{2004}{185504}.


 \Name{Marinari E.\and Pitard E.}
  \REVIEW{Europhys. Lett.}{69}{2005}{235}.


\bibitem{JKGC} 
 \Name{Jack R.,Kelsey D., Garrahan J.P. \and Chandler D.}
  \REVIEW{Phys. Rev. E}{78}{2008}{11506}.


\bibitem{Giardina}
 \Name{Giardin\`a C., Kurchan J. \and Peliti L.}
  \REVIEW{Phys. Rev. Lett.}{96}{2006}{120603}.


\bibitem{Bodineau-Derrida}
 \Name{Bodineau T. \and Derrida B.}
  \REVIEW{Phys. Rev. E}{72}{2005}{066110}.

  
   \Name{Lecomte V.}
  \REVIEW{PhD thesis}{2007}.
  
\bibitem{Lebowitz-Spohn} 
 \Name{Lebowitz J.L.\and Spohn H.}
  \REVIEW{J. Stat. Phys.}{95}{1999}{333}.



\bibitem{dauchot-levy} 
 \Name{Lechenault F., Dauchot O., Biroli G. \and Bouchaud J.P.}
  \REVIEW{Europhys. Lett}{83}{2008}{46003}.




\bibitem{biggerpaper} 
 \Name{Turci F.\and Pitard E.} in preparation.

\bibitem{Speck-Garrahan} 
 \Name{Speck T.\and Garrahan J.P.} arXiv:1004.2698.








\end{thebibliography}
\end{document}